\documentclass[%
 aip,
 amsmath,amssymb,
 reprint,%
]{revtex4-1}

\usepackage{graphicx}
\usepackage{dcolumn}
\usepackage{bm}
\usepackage{float} 

\usepackage[utf8]{inputenc}
\usepackage[T1]{fontenc}
\usepackage{mathptmx}
\usepackage{etoolbox}
\usepackage{color,soul} 
\usepackage{xcolor} 
\usepackage[colorlinks=true, linkcolor=blue, citecolor=blue, urlcolor=blue]{hyperref}

\makeatletter
\def\@email#1#2{%
 \endgroup
 \patchcmd{\titleblock@produce}
  {\frontmatter@RRAPformat}
  {\frontmatter@RRAPformat{\produce@RRAP{*#1\href{mailto:#2}{#2}}}\frontmatter@RRAPformat}
  {}{}
}%
\makeatother
\begin{document}

\preprint{AIP/123-QED}

\title[]{Efficient detection of spectral cluster states}
\author{M. Kalash}
 
\affiliation{ 
Friedrich-Alexander Universit\"at Erlangen-N\"urnberg, Staudtstra\ss{}e 7/B2, 91058 Erlangen, Germany
}
\affiliation{Max Planck Institute for the Science of Light, Staudtstra\ss{}e 2, 91058 Erlangen, Germany}
\author{U.-N. Han}%
\affiliation{%
Department of Physics, Korea Advanced Institute of Science and Technology (KAIST), Daejeon 34141, Korea}%

\author{C. Roh}%
\affiliation{%
Department of Physics, Korea Advanced Institute of Science and Technology (KAIST), Daejeon 34141, Korea}%
\author{M. H. M. Passos}
 \affiliation{Max Planck Institute for the Science of Light, Staudtstra\ss{}e 2, 91058 Erlangen, Germany}
 \affiliation{Centro Brasileiro de Pesquisas Físicas, Rua Dr. Xavier Sigaud, 150, Urca, Rio de Janeiro, 22290-180, RJ, Brazil}
\author{Y.-S. Ra}
\affiliation{%
Department of Physics, Korea Advanced Institute of Science and Technology (KAIST), Daejeon 34141, Korea}%
\author{M. V. Chekhova}

\affiliation{ 
Friedrich-Alexander Universit\"at Erlangen-N\"urnberg, Staudtstra\ss{}e 7/B2, 91058 Erlangen, Germany
}%
\affiliation{Max Planck Institute for the Science of Light, Staudtstra\ss{}e 2, 91058 Erlangen, Germany}
 \email{maria.chekhova@mpl.mpg.de}

\date{\today}

\begin{abstract}
Multimode entangled states of light are key resources for
high-dimensional photonic quantum technologies, particularly for
continuous-variable quantum computing and quantum information processing. Exploiting their
full capabilities requires detection strategies that can efficiently
and simultaneously access a large number of modes and their quantum
correlations. Multimode optical parametric amplification (MOPA)
provides a promising approach by amplifying multiple quadratures
prior to detection, rendering the measurement robust against
subsequent optical losses while enabling broadband operation. Here,
we demonstrate the spectral multimode capability of MOPA by
simultaneously characterizing a broadband 60-node dual-rail
continuous-variable cluster state. We reconstruct the cluster-state nullifiers across the
spectral modes and observe nearly uniform squeezing of
up to $-6.9$ dB, {which, to the best of our knowledge,
is the highest level reported for pulsed spectrally
multimode squeezed light.} Furthermore, by evaluating
the van Loock--Furusawa criterion across all relevant bipartitions,
we verify genuine multipartite entanglement throughout the 60-node
state. These results  establish MOPA as a
promising platform for scalable high-dimensional quantum information
processing in the spectral domain.
\end{abstract}

\maketitle

\section{\label{intro}Introduction}
A central resource for measurement-based quantum computation (MBQC)
is the continuous-variable (CV) cluster state, which serves as the
computational {base} of the protocol
\cite{Raussendorf2003Aug,Menicucci2006Sep}. Optical cluster states
are multipartite entangled networks whose nodes encode quantum
information in the quadratures of light
\cite{Menicucci2011Apr}. Their deterministic generation from
multimode squeezed light using linear-optical networks makes them
particularly attractive for photonic quantum information processing
\cite{vanLoock2007Sep}. Exploiting the full capabilities of these
states, however, requires detection strategies that simultaneously
access all constituent nodes and links, enabling parallel quantum
operations. Although homodyne detection can provide such multimode access
through dispersive and mode-sorting elements combined with array
detectors~\cite{cai2017,RomanRodriguez2024Nov,Armstrong2012Aug,yoon2026}, these additional optical components introduce losses.
Moreover, current detector technologies impose trade-offs between
detection efficiency, bandwidth, and acquisition speed, ultimately
limiting the scalability of multimode implementations.

Multimode optical parametric amplification (MOPA)
offers an alternative detection paradigm. By simultaneously amplifying all optical modes prior to measurement,
MOPA removes the effect of subsequent losses, thereby
enabling the efficient use of mode demultiplexers and detector arrays~\cite{kalash2026}.
Furthermore, because
OPAs are the inherent broadband and multimode sources of squeezed light,
they are naturally suited to detecting such high-dimensional states. Therefore, the accessible optical bandwidth is not
restricted by that of a local oscillator, which is typically narrower
than the bandwidth of the states themselves. Finally, because the
measurement is performed optically within the nonlinear interaction,
it bypasses electronic bandwidth limitations~\cite{Shaked2018Feb} and opens a route towards
ultrafast multimode detection. 

Previous work has already demonstrated the potential of this
approach for efficient squeezing detection. Optical parametric
amplification has enabled the measurement of spectrally~\cite{Shaked2018Feb,Takanashi2020Nov} and spatially
broadband squeezing~\cite{Frascella2019Oct}, few-cycle squeezing~\cite{nehra2022a},
simultaneous detection of multiple squeezed modes~\cite{Barakat2025Feb}, and,
ultimately, the real-time access to squeezing and entanglement
across many co-propagating spatial modes~\cite{kalash2026}.

In this work, we explore, for the first time, the spectral multimode capabilities of
MOPA and demonstrate them experimentally by simultaneously
characterizing a high-quality broadband 60-node optical cluster
state. {We measure cluster correlations as strong as
$-6.9\,\mathrm{dB}$ below the shot-noise limit, with
a nearly constant nullifier-mode purity of $73\%$
across all modes.} Furthermore, we verify genuine
multipartite entanglement by analyzing the inseparability across all
nodes. These results establish MOPA as a promising platform for the
detection of spectrally multipartite entangled states and open new
opportunities for scalable quantum technologies based on optical
frequency combs and integrated photonic circuits~\cite{chang2025,kues2019,roh2025,yang2021,zhao2020,guidry2023,jahanbozorgi2023,jia2025a}.

\begin{figure*}[t!]
\centering
\includegraphics[width=1\linewidth]{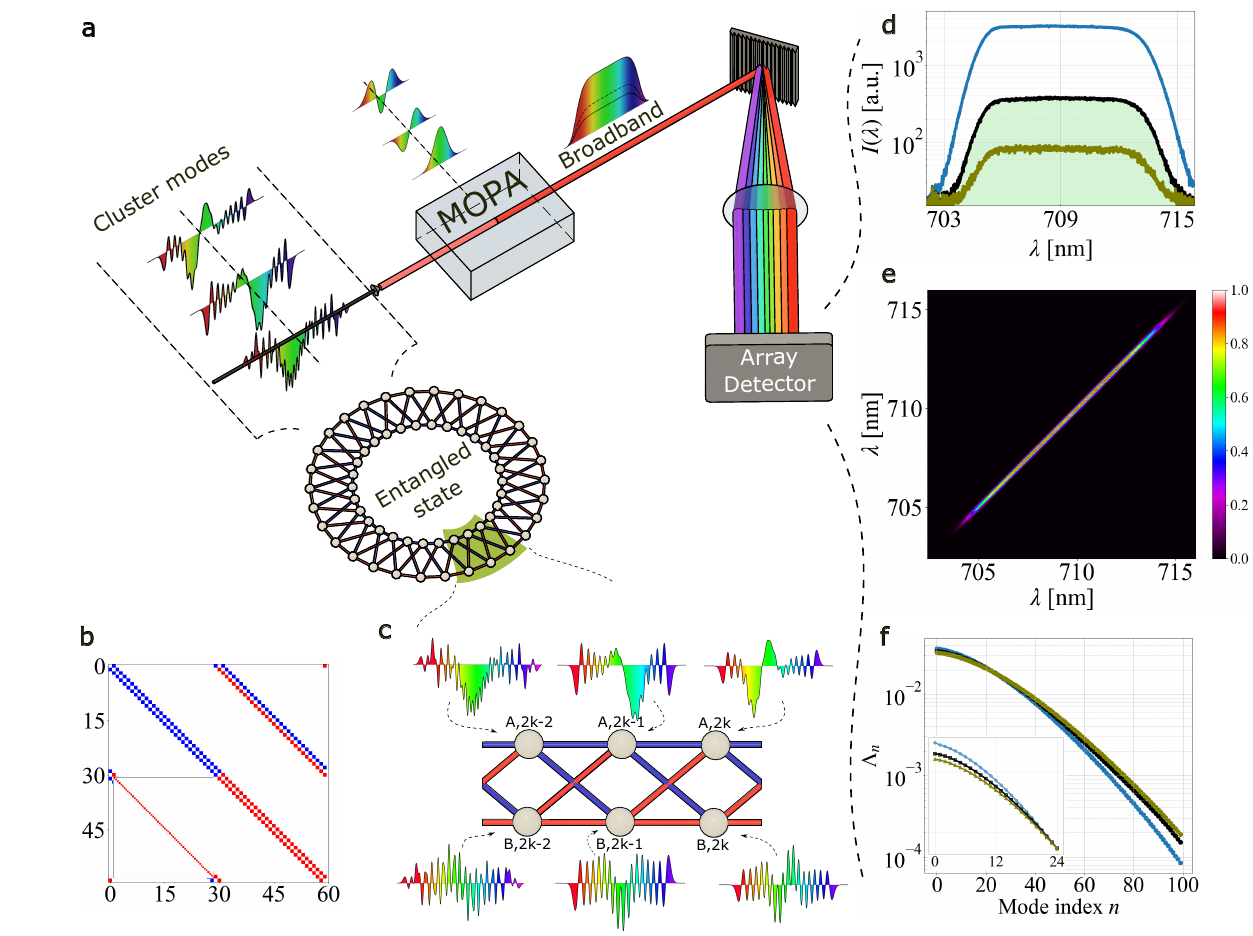}
     \caption{Simultaneous characterization of spectral cluster states
through multimode optical parametric amplification (MOPA).
(a) The target state, here a 60-node dual-rail cluster
state, is injected into the MOPA, which performs broadband
phase-sensitive amplification. The amplified radiation is then
spectrally dispersed by a diffraction grating and measured with an
array detector, providing access to the spectral intensity and its
correlations. (b) Adjacency matrix of the target 60-node dual-rail
cluster state, with its square shown in the inset. Blue and red points correspond to connections of weights $-1/2$ and $1/2$, respectively. (c) Spectral profiles of the cluster modes corresponding
to some of the nodes. (d) Measured spectra {at the MOPA output for  vacuum input (black) and for squeezed vacuum input with}
phases $\phi=0$ (blue) and $\phi=\pi/2$ (olive). The green
region below the amplified-vacuum spectrum indicates squeezing. (e) Measured spectral intensity correlations of the
amplified vacuum. (f) Reconstructed weights of the 100 strongest modes for
amplified vacuum (black), $\phi=0$ (blue), and $\phi=\pi/2$ (olive).} 
    \label{scheme}
\end{figure*}

\section{\label{Results}Results}
\subsection*{Source}
For the generation of the cluster state, we employ multimode
squeezing as the entanglement resource~\cite{braunstein2005a}. To this end, type-I
collinear degenerate parametric down-conversion (PDC) is generated in a
3 mm thick bismuth triborate (BiBO) crystal pumped by an 18 ps,
1 kHz laser at 354.67 nm. The modal content of the resulting
multimode squeezed state is determined by the pump spectral
profile, the phase-matching function of the nonlinear crystal, and
the parametric gain~\cite{christ2013,scharwald2026}. 
{For a gain of $G_{sq}=1.18\pm0.05$, the generated state
is expected to contain a large number of squeezed spectral
modes spanning a bandwidth of 37 nm, with profiles closely
following Hermite--Gaussian distributions and a fundamental
mode width of 3.4 nm (see Supplementary Information).}

From the available multimode squeezing, different entangled states
can be generated through suitable mode transformations~\cite{vanLoock2007Sep}. The modal
structure of each state is determined by how the initial squeezed
modes are combined~\cite{ferrini2015a}. For example, an
Einstein--Podolsky--Rosen (EPR) pair is obtained by mixing two
independently squeezed modes on a beam splitter. It is characterized
by the quadrature combinations
$\hat{\delta}_x=\hat{x}_A-\hat{x}_B$ and
$\hat{\delta}_p=\hat{p}_A+\hat{p}_B$, called nullifiers, whose simultaneous squeezing
indicates entanglement.
\begin{figure*}[t!]
\centering
\includegraphics[width=0.75\linewidth]{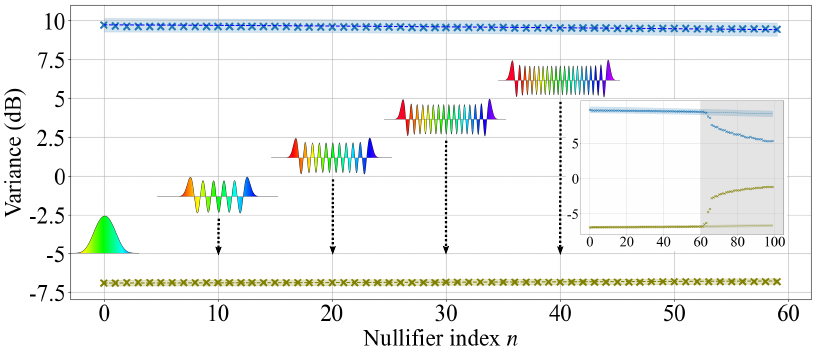}
\caption{Measured nullifier variances of the 60-node dual-rail cluster state. {Crosses show reconstructed} variances at relative phases
$\phi=0$ (blue) and $\phi=\pi/2$ (olive). The dashed lines show the theoretical values, while the shaded  areas account for the uncertainty in the initial squeezing level. The {colored shapes are the} spectral profiles of 
the 1st, 10th, 20th, 30th, and 40th nullifiers. The inset
shows the reconstructed quadrature variances of the 100 strongest
{frequency modes of the squeezed vacuum}.} 
    \label{squeezingmodes}
\end{figure*}

More complex states can be generated by
combining a larger number of squeezed modes~\cite{braunstein2005a}. Their structure is
conveniently represented by an adjacency matrix $V$~\cite{gu2009a}, which specifies
the connections between the constituent nodes. 
Equivalently, rather than physically transforming the squeezed modes,
the same entanglement structure can be accessed by measuring the
multimode squeezed state in an appropriate modal basis~\cite{Armstrong2012Aug}. Different
cluster states can therefore be realized and characterized simply by
changing the measurement basis~\cite{cai2017}. Unlike for an EPR pair, verifying
entanglement in a multipartite state requires testing suitable
quadrature combinations across all relevant bipartitions~\cite{vanLoockFurusawaCrt}.

\subsection*{Target state}
{In our proof-of-principle experiment, we focus on the
dual-rail CV cluster state
~\cite{Menicucci2011,Yokoyama2013Dec,Larsen2019Oct,
Asavanant2019Oct}, a scalable resource for measurement-based
quantum computation. In higher-dimensional configurations,
this resource also enables deterministic and programmable
multimode Gaussian operations~\cite{Larsen2021}.} The graph representation of the target 60-node state is presented in Fig. \ref{scheme}(a), while its adjacency matrix $V$ is shown in Fig. \ref{scheme}(b). Starting from the squeezed modes, the transformation defined by
$V$ yields the cluster-state modes, whose spectral profiles for several nodes are shown
in panel~(c). A key feature of this
graph is that it is self-inverse, i.e. $V^2=\mathbb{I}$, as illustrated
in the inset. Consequently, the cluster-state nullifier {modes are orthogonal}, and they coincide with the modal
basis of the underlying multimode squeezed state (see Supplementary Information).
Dual-rail cluster states can be viewed as extensions of EPR states up to local phase-space rotations (by $\pi/2$) \cite{Menicucci2011,Asavanant2019Oct}. Accordingly, the corresponding nullifiers generalizing the bipartite
EPR correlations to the multimode cluster-state setting are given
by
\begin{equation}
\begin{aligned}
\hat{\delta}_{A,2k-1}
&=
\hat{P}_{A,2k-1}
+ \tfrac{1}{2}\!\left(
\hat{X}_{A,2k-2}
+ \hat{X}_{A,2k}
- \hat{X}_{B,2k-2}
+ \hat{X}_{B,2k}\right),\\[5pt]
\hat{\delta}_{B,2k-1}
&=
\hat{P}_{B,2k-1}
+ \tfrac{1}{2}\!\left(
\hat{X}_{A,2k-2}
- \hat{X}_{A,2k}
- \hat{X}_{B,2k-2}
- \hat{X}_{B,2k}\right),
\end{aligned}
\end{equation}
where the node indices are defined in terms of $k$, and $A$ and $B$ denote the upper and lower rails, respectively \cite{Asavanant2019Oct,Chen2014Mar} (see Supplementary Information).

\subsection*{Multimode optical parametric amplification}
Multimode optical parametric amplification (MOPA) is
implemented using the same PDC process employed for generating the squeezed
state, but now operated at
a higher gain \(G=4.5\pm0.2\). In this regime, the MOPA functions as an
all-optical analyzer of an input multimode squeezed state, transforming the
fragile quadratures fluctuations into macroscopic intensity
observable that can be measured directly with conventional
photodetectors~\cite{Shaked2018Feb}.

Meanwhile, the measurement basis {can be configured by tailoring the MOPA pump
profile, which controls} the eigenmodes of the amplifier~\cite{kalash2026}.
These eigenmodes can therefore be matched to a specific set of
target modes. Since the correlations of the target cluster state are encoded in the modal basis of the underlying multimode squeezed state, the same
pump profile used for squeezing can be retained for the MOPA (see Supplementary Information).
{The amplifier eigenmodes therefore follow the same
Hermite--Gaussian basis, but differ in spectral width
due to gain-induced broadening~\cite{scharwald2026}. Indeed, because $G>G_{sq}$, the amplifier modes are
broader than the squeezed ones, with the calculated
fundamental-mode bandwidth increasing to 4.9 nm.}
\begin{figure*}[t!]
\centering
\includegraphics[width=0.7\linewidth]{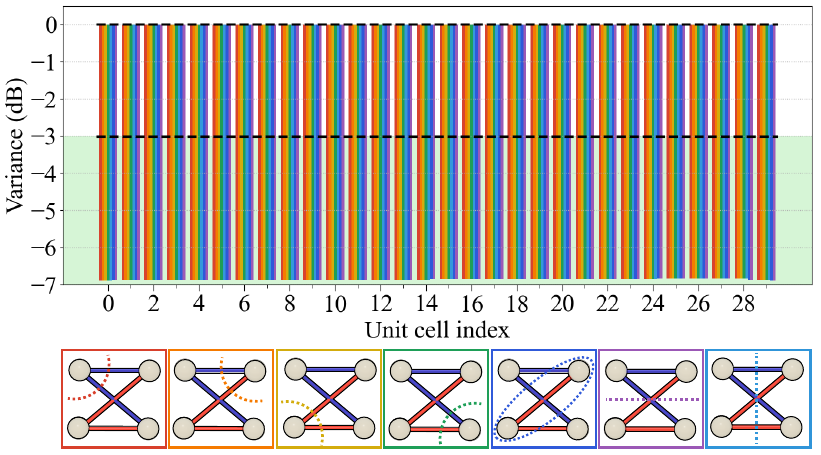}
\caption{Verification of multipartite entanglement. The vLF criterion is evaluated for all 30 unit cells
comprising the 60-node dual-rail cluster state. The seven possible
bipartitions of the chosen unit cell configuration are illustrated below the main panel,
with each identified by a different color. Each bar shows the measured vLF value for the corresponding unit cell
and the bipartition identified by its matching color. The horizontal dashed lines at $0$ dB and $-3$ dB indicate
the vLF bounds for the far-right bipartition (light blue) and all
other bipartitions, respectively. Values below the corresponding
bounds verify inseparability across each bipartition.} 
    \label{VLF}
\end{figure*}

Nevertheless, perfect mode matching between the input nullifier modes
and the MOPA eigenmodes is not required for squeezing detection.
Information carried by an input mode is generally distributed among
several amplifier eigenmodes according to the overlaps $g_{ln}$ (see Supplementary Information).
Consequently, the input quadrature variance can be fully recovered
from a weighted combination of the measured intensities in all output
channels as~\cite{Barakat2025Feb}
\begin{equation}
\operatorname{Var}\!\left(\delta_l^\phi\right)
\propto
\sum_n g_{ln}^{\,2}
\mathrm{e}^{-2G_n}
I_n^\phi(\lambda),
\label{sqnull}
\end{equation} where $\delta_l^\phi$ denotes the $l$th generalized nullifier,
$G_n$ is the parametric gain of the $n$th amplifier eigenmode, and
$I_n^\phi(\lambda)$ is the corresponding phase-sensitive output
intensity measured in the $n$th MOPA output channel.
This retrieval is exact provided that the overlap matrix $g_{ln}$ is
real, which requires the relative phases acquired between the squeezer
and the MOPA to be negligible or compensated. For more general cases
involving complex overlaps, the MOPA eigenmodes must be appropriately
engineered.

\subsection*{Simultaneous characterization of spectral dual-rail cluster state}
We now apply MOPA to the simultaneous characterization
of the 60-node cluster state, as illustrated in
Fig.~\ref{scheme}(a). Through broadband phase-sensitive
amplification (PSA), the entire spectrum $I^\phi (\lambda)$ is enhanced or suppressed
relative to the amplified-vacuum reference (dashed line),
depending on the phase $\phi$. The amplified radiation is then
dispersed by a diffraction grating, providing access to both
the spectral intensity and its correlations using an array detector. From the latter,
the spectral modes and their corresponding nullifier variances
can be reconstructed~\cite{finger2017}. The full experimental setup is presented in the supplementary document.

Figure~\ref{scheme}(d) shows the measured amplified spectrum $I^\phi (\lambda)$.
To ensure a uniform amplification phase, we exclude spectral regions
affected by dispersion (see Supplementary Information) by restricting
the measurement bandwidth to $8.6\,\mathrm{nm}$ using a bandpass
filter, although a suitable filter would allow extending the
bandwidth to approximately $20\,\mathrm{nm}$ while maintaining the
uniform phase. The flat-top amplification profile is evident from the
amplified-vacuum spectrum shown in black. After PSA, spectral components falling below the
amplified-vacuum reference reveal squeezing in the amplified
quadratures, as indicated by the green region in
Fig.~\ref{scheme}(d). By setting $\phi=\pi/2$, all squeezed
quadratures are amplified uniformly, as shown by the olive solid
line, resulting in squeezing across the entire recorded spectrum. Similarly, setting $\phi=0$ uniformly amplifies all anti-squeezed
quadratures, resulting in intensities above the amplified-vacuum reference across the recorded spectrum (blue line). 

Having obtained the broadband PSA spectrum, we
perform coherent mode decomposition through the spectral intensity
covariance function
$\operatorname{Cov}(\lambda,\lambda')
=\langle I^\phi(\lambda)I^\phi(\lambda')\rangle
-\langle I^\phi(\lambda)\rangle
\langle I^\phi(\lambda')\rangle$.
As shown in reference~\cite{Barakat2025Feb}, its singular-value decomposition yields the modal weights $\Lambda_n^\phi$, which
quantify the contribution of each mode $I_n^\phi(\lambda)$ to $I^\phi(\lambda)$.
The corresponding modal intensities can then
be obtained and used in Eq.~\eqref{sqnull} to reconstruct the target nullifier
variances. Figure~\ref{scheme}(e) shows, as an example, the measured spectral
covariance function for the amplified vacuum. From this covariance
and those measured at $\phi=0$ and $\phi=\pi/2$, the corresponding
modal weights $\Lambda_n^\phi$ are reconstructed, with the first
100 shown in panel~(f), with this number being primarily limited by the applied
spectral filtering (see Supplementary Information).
Further, accounting for intermodal coupling through
Eq.~\eqref{sqnull} and the effects of spectral filtering
(see Supplementary Information), we simultaneously reconstruct the squeezing levels of 100 spectral modes, as shown in the inset of Fig.~\ref{squeezingmodes}, including the 60 modes corresponding to the cluster-state nullifiers (Fig.~\ref{squeezingmodes} main panel). Olive (blue) crosses show the experimentally reconstructed squeezing (anti-squeezing) levels, while the dashed lines and shaded areas represent the theoretical prediction. The field distributions of the 1st, 10th, 20th, 30th, and 40th modes are shown as insets. Nearly uniform squeezing is observed across all 60
cluster-state nullifiers, with squeezing levels ranging
from $-6.9$ to $-6.8$~dB, demonstrating both the
high quality of the characterized cluster state and the
highly multimode detection capability of MOPA.
For all nullifiers, the purity remains nearly constant at $73\%$.
Although squeezing can be reconstructed for up to 100
spectral modes, agreement with the theoretical predictions
is limited to the first 60. This is because correcting
for intermodal coupling requires accounting for
higher-order MOPA modes: the 60th squeezed mode already
couples to the 100th MOPA mode, while higher-order
squeezed modes couple to MOPA modes inaccessible within
the detection bandwidth. Without the $g_{ln}$ correction,
nearly uniform squeezing is observed across all 100
spectral modes, corresponding to superpositions of
the input squeezed modes rather than the individual
cluster-state nullifiers (see Supplementary Information). The number of accessible
modes could be further increased by extending the
detection bandwidth. To the best of our knowledge, this represents the
highest squeezing level reported to date for pulsed {spectrally} multimode squeezed light.

Finally, to verify genuine multipartite entanglement across the
target state, we evaluate the van Loock--Furusawa (vLF) criterion
\cite{vanLoockFurusawaCrt} (see Supplementary Information) over 30 unit cells, each comprising four
neighboring nodes, as illustrated in Fig.~\ref{VLF}. For the chosen unit-cell configuration, the vLF criterion is
evaluated over seven distinct bipartitions (Fig.~\ref{VLF}, bottom panels). Each bipartition is identified by a different color,
while the dashed lines indicate the division of the four nodes into
the two subsystems considered in the vLF criterion. This evaluation
is repeated for all 30 unit cells. All bipartitions have a vLF bound
of $-3$ dB, except for the vertical bipartition (light blue), for
which the bound is $0$ dB. As shown, all measured values lie below their respective vLF bounds,
demonstrating inseparability for every bipartition and thereby
confirming genuine multipartite entanglement across all 60 spectral
modes. 
\section{Discussion}
In summary, we have 
{characterized} a 60-node
dual-rail spectral cluster state using multimode optical parametric
amplification (MOPA) followed by direct detection. In our approach, MOPA
maps the variances of the cluster-state nullifiers onto measurable
{intensities}, enabling their simultaneous direct loss-tolerant detection with a
camera. Importantly, nullifier squeezing can be retrieved even without
perfect mode matching between the cluster-state nullifiers and the
MOPA eigenmodes, provided that the relative phases between the
input and MOPA modes are uniform. For more complex modal structures, the MOPA eigenmodes must be
appropriately shaped, which can be achieved using standard
pulse-shaping techniques applied to the MOPA pump.

Among hundreds of spectral modes, we observed nearly uniform
squeezing ranging from $-6.9$ to $-6.8$ dB across 60 {nullifier modes, the purity being $73\%$}. To the best of our knowledge,
this is the highest squeezing level reported for pulsed {spectrally}
multimode squeezed light. The violation of the vLF bounds across all relevant bipartitions
further confirms genuine multipartite entanglement across the
60-node state. The number of
accessible modes was primarily limited by spectral dispersion, which
necessitated spectral filtering to maintain a uniform phase across
the recorded spectrum, restricting the measurement bandwidth to
$8.6\,\mathrm{nm}$. We expect that dispersion compensation will
eliminate the need for such filtering, enabling access to a broader
spectral range and, consequently, a larger number of modes.

Our work expands the scope of MOPA-based quadrature detection from
spatial~\cite{Barakat2025Feb,kalash2026} to spectral modes, and from squeezing detection~\cite{Shaked2018Feb,Takanashi2020Nov,nehra2022a,Frascella2019Oct} to
entanglement characterization, demonstrating the broad versatility
of the approach. It also highlights MOPA's ability to simultaneously access a large
number of modes, offering a significant advantage for
high-dimensional continuous-variable quantum information processing,
particularly in integrated photonic platforms operating in the
spectral domain. While the modal information is reconstructed here
through spectral correlation analysis in post-processing, combining
MOPA with spectral mode-sorting techniques, such as quantum pulse
gates~\cite{serino2023a}, could enable simultaneous direct access to the individual
modes, opening a route towards real-time monitoring of multipartite
entanglement.

\section*{Acknowledgments}
M.K. and M. V. C. acknowledge funding from European Research Council (Project 101199215 — MultiFlaQS) and from Max-Planck School of Photonics. U.-N. Han acknowledges funding from the Ministry of Science and ICT (MSIT) of Korea (2021M3H3A1036573). U.-N. Han and Y.-S. Ra acknowledge funding from the MSIT of Korea (RS-2024-00442762 and RS-2025-25464959).
\section*{Data availability }
The data supporting the findings of this study are available from the corresponding author upon request.
\section*{Author contributions}
M.K. and M.V.C. conceived the experiment. M.K. and U.-N.H. performed the experiment and analysed the data. C.R. and M.H.M.P. contributed to the theory. U.-N.H. prepared the figures and wrote the Supplementary Information. M.K., M.H.M.P. and M.V.C. wrote the manuscript. M.V.C. and Y.-S.R. supervised the project. All authors discussed the results and reviewed the manuscript.
\section*{Conflict of interest}
The authors declare no competing interests.
\bibliography{Modes}

\end{document}